\newcommand{\NB}{{\it NB816}}
\documentclass[12pt,preprint]{aastex}

\shortauthors{Ajiki et al.}
\shorttitle{LYMAN$\alpha$ EMITTERS AT $z = 5.7$}

\begin{document}

% \submitted{}

\title{Narrow-band Survey of the GOODS Fields: Search for LYMAN$\alpha$ Emitters at $z=5.7$ \footnotemark[1]}

\author{Masaru Ajiki\altaffilmark{2},
 Bahram Mobasher\altaffilmark{3,4},
 Yoshiaki Taniguchi\altaffilmark{2},
 Yasuhiro Shioya\altaffilmark{2},
 Tohru Nagao\altaffilmark{5,6},
 Takashi Murayama\altaffilmark{2}, and
 Shunji S. Sasaki\altaffilmark{2}
}

\footnotetext[1]{Based in part on data collected at Subaru Telescope and 
obtained from the SMOKA science archive at Astronomical Data Analysis Center, 
which are operated by the National Astronomical Observatory of Japan.}

\altaffiltext{2}{Astronomical Institute, Graduate School of Science,
 Tohoku University, Aramaki, Aoba, Sendai 980-8578, Japan}
\altaffiltext{3}{Space Telescope Science Institute 3700 San Martin Drive Baltimore MD 21218}
\altaffiltext{4}{Also affiliated with the Space Sciences Department of the European Space Agency}
\altaffiltext{5}{National Astronomical Observatory, Mitaka, Tokyo 181-8588, Japan}
\altaffiltext{6}{INAF --- Osservatorio Astrofisico di Arcetri, Largo Enrico Fermi 5, 50125 
                 Firenze, Italy}

\begin{abstract}

We present results from optical narrow-band ($\lambda_{\rm c} = 8150$ \AA ~ and $\Delta\lambda = 120$ \AA)
observations of the Great Observatories Origins Deep Survey (GOODS) fields, 
using Suprime-Cam on the Subaru Telescope. Using these narrow-band data, we
then perform a survey of Lyman $\alpha$ Emitters (LAEs) at $z\sim 5.7$. 
The LAE survey covers an area of $\approx 320$ arcmin$^2$ and a co-moving 
volume of $\simeq 8.0 \times 10^4$ Mpc$^3$. We found a total of 10 (GOODS-N)
and 4 (GOODS-S) LAE candidates at $z\sim 5.7$. We perform a study of the 
spatial distribution, space density, and star formation properties of the LAEs
at $z\sim 5.7$.

\end{abstract}


\section{INTRODUCTION}

With the installation of the Advanced Camera for Surveys (ACS), the
 efficiency of the {\it Hubble Space Telescope (HST)} in performing wide-area
 and deep surveys has increased ten-fold. Such data are required to study 
 formation and evolution of galaxies and are essential
 in identifying high-redshift galaxies.
Significant progress has been made in recent years in
performing multi-waveband deep surveys, greatly extending the range of
accessible redshifts and enhancing our knowledge of properties of galaxies
out to the re-ionization epoch (Beckwith et al. 2004; Giavalisco et al. 2004a; Rix et al. 2004; 
 Scoville et al. 2006 in preparation).   

One of the general methods to select galaxies at high redshift is
drop-out technique (e.g., Steidel et al. 1996), monitoring the Lyman break (912 \AA)
feature in galaxies as it moves with redshift to longer wavelengths and 
progressively out of the observed (shorter) passbands.
Such high-z galaxies have been found in large numbers in the Great Observatories
Origins Deep Survey fields (GOODS; Giavalisco et al. 2004b; 
Dickinson et al. 2004; Stanway et al. 2004; Bouwens et al. 2004) and
are used to investigate the origin of cosmic re-ionization at $z \gtrsim 6$ 
(e.g., Barkana \& Loeb 2001).
 Another technique to identify high-$z$ objects is narrow-band 
search for strong Lyman $\alpha$ emitters (LAEs)
(e.g., Hu, Cowie \& McMahon 1998; Cowie \& Hu 1998; Rhoads \& Malhotra 2001;
 Hu et al. 2002; Ajiki et al. 2003; Kodaira et al. 2003, Taniguchi et al. 2005).
Such narrow-band surveys, conducted at selected wavelengths to avoid
 atmospheric emission, target emission line galaxies at certain redshifts. 
Combined with broad-band data to subtract the continuum, we can estimate the 
narrow-band flux for the respective lines observed. 

In this paper, we present results from a narrow-band survey in the 
GOODS fields, searching for LAEs at $z=5.7$. The area coverage and depth 
of the GOODS, combined with its 
multi-wavelength imaging data, allows unbiased selection of a population of
LAEs at high redshifts. Furthermore, 
availability of deep Spitzer and Chandra data, high resolution ACS images and 
ground-based optical-IR data in the GOODS fields allow a study of the nature
of LAEs, their morphology and star formation rate at $z=5.7$  and provides an
unbiased sample of LAEs to compare with homogeneously selected LBGs at 
a similar redshift.  
We adopt a flat universe with $\Omega_{\rm matter} = 0.3$,
$\Omega_{\Lambda} = 0.7$, and $H_0 = $70 km
s$^{-1}$ Mpc$^{-1}$. Throughout this paper,
magnitudes are given in the AB system.

\section{OBSERVATIONS AND DATA REDUCTION}

\subsection{Narrow-band Observations and Source Detection}

Deep narrow-band surveys of the GOODS fields are performed 
using the Suprime-Cam (Miyazaki et al. 2002) on the 8.2 m Subaru Telescope. 
We use a narrowband filter, \NB, centered on 8150 \AA ~ with a width 
$\Delta\lambda_{\rm FWHM} = 120$ \AA (Ajiki et al. 2003). This filter
samples Lyman $\alpha$ emission lines in the redshift range $z=5.65-5.75$. 

The \NB ~ observations of GOODS-S were made during 20--22 February 2004 (UT) 
while, for the GOODS-N field, we used Subaru archival data which were taken 
in April 2002 and 2003.  All the \NB ~ observations were done under photometric
conditions with a seeing between 0$\farcs$7 and 0$\farcs$9.  
The narrowband imaging data were reduced
with the software package SDFRED (Yagi et al. 2002; Ouchi et al. 2004).
The FWHM of stellar objects in the final {\it NB816} images of the two fields are 0$\farcs$9.
The limiting \NB ~ magnitude in both fields is $\NB \approx25.4$ 
for a 5$\sigma$ detection over a $2^{\prime \prime}$ diameter aperture.  
Deep broad-band images taken by the ACS ($B_{435}$, $V_{606}$, $i_{775}$,
and $z_{850}$) and ground-based near-IR data, taken as a part of the GOODS 
(Giavalisco et al. 2004a) are also available for these fields.

The \NB ~ survey covers a total of 160 arcmin$^2$ area in each field, 
corresponding to the
full GOODS-ACS area (a total of 320 arcmin$^2$ area). The transverse 
co-moving area of the LAE survey at $z=5.7$ is 
1.8 $\times 10^{3}$ Mpc$^2$. Combined with the FWHM of the
filter, which has a Gaussian-like shape, this corresponds to a co-moving 
depth of 45 Mpc ($z_{\rm min} \approx 5.65$
to $z_{\rm max} \approx 5.75$) and a combine volume of
$8.0 \times 10^{4}$ Mpc$^{3}$. 

Source detection was performed using SExtractor version 2.3 
(Bertin \& Arnouts 1996).  
A source is defined as 9-pixel contiguous area above the 3 $\sigma$ noise level
on the \NB ~ image. Photometry was performed using IRAF -- APPHOT with a 
$2^{\prime \prime}$ diameter aperture.  
To the magnitude limit of our survey, $\NB = 25.4$
 ($\approx 5\sigma$ in the both fields), a total of 9,600
 and 8,700 sources are found in the GOODS-N and -S fields respectively.

\subsection{Selection of LAE Candidates}

\begin{figure}
\epsscale{1.0}
\plottwo{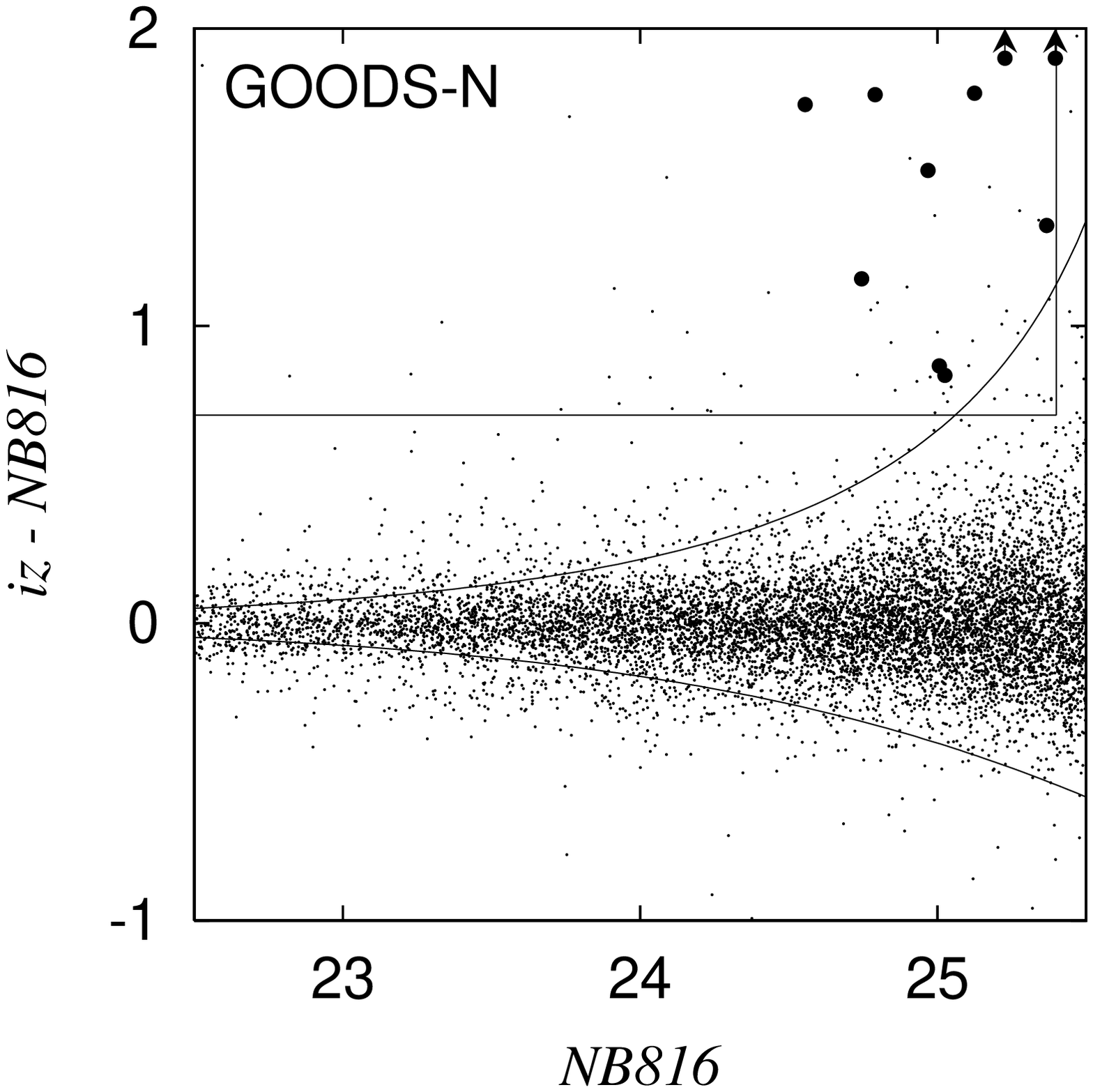}{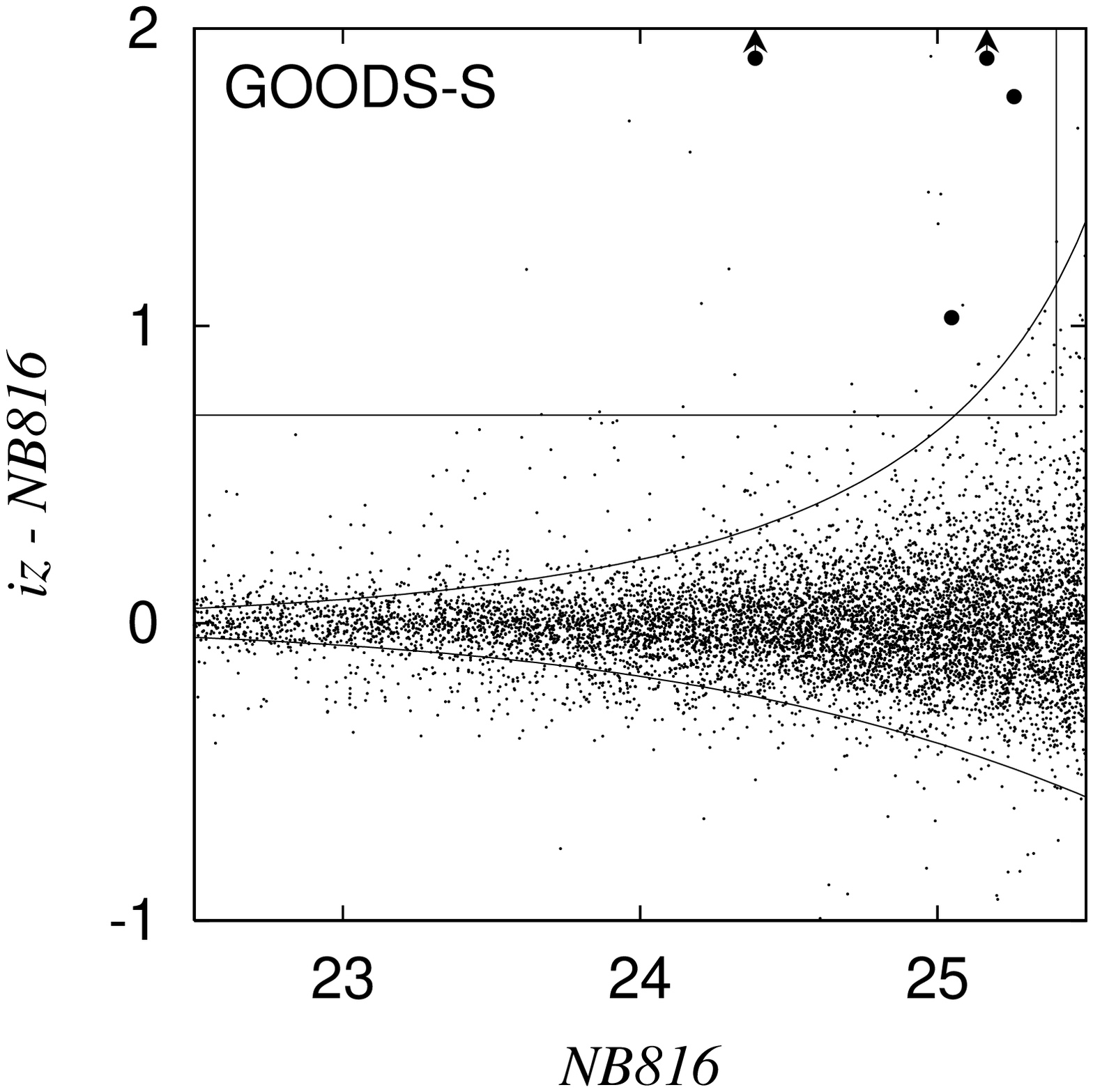}
\caption{$iz-\NB$  vs. \NB ~ diagram in GOODS-N (left) 
         and GOODS-S (right).
         Our LAE candidates are shown as filled circles.
         The vertical lines show the detection limit corresponding to  \NB$=25.4$.
         The horizontal lines show the selection criterion $iz- \NB= 0.7$.
         The curves show the 3$\sigma$ limits for $iz-\NB$.
\label{nbcm}}
\end{figure}

\begin{figure}
\epsscale{1.0}
\plotone{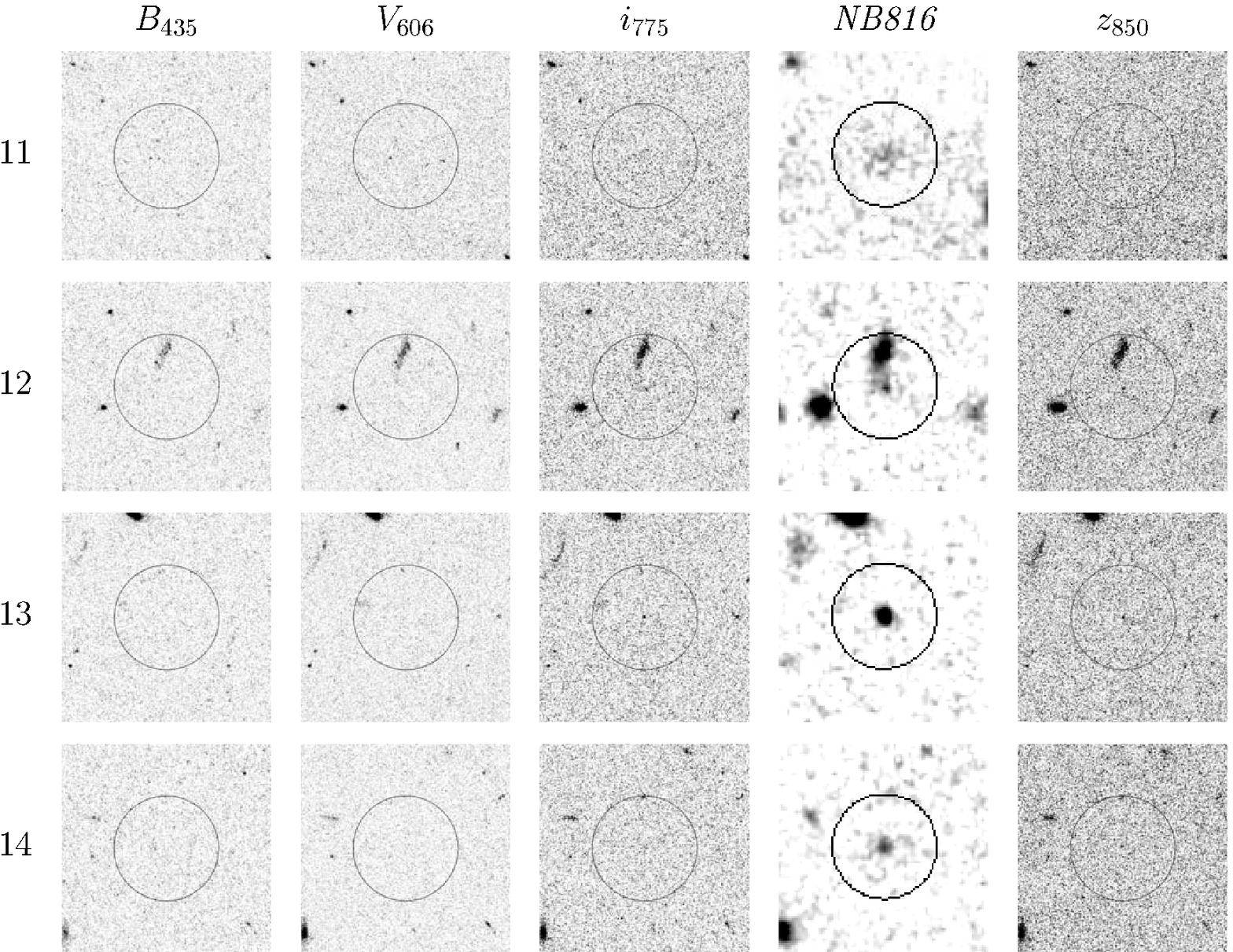}
\caption{Broad-band and \NB ~ images of our
          LAE candidates at $z \approx 5.7$ GOODS-S.
         Each box is $12^{\prime \prime}$ on a side
         (north is up and east is left).
         Each circle has $3^{\prime \prime}$ radius.
\label{thumS}}
\end{figure}

The LAE candidates at $z \approx 5.7$ are identified by combining the
GOODS ACS images ($B_{435}$, $V_{606}$, $i_{775}$, and $z_{850}$) 
(Giavalisco et al. 2004a) and narrow-band catalog with $\NB<25.4$.
The LAEs are selected to satisfy the following criteria:
\begin{eqnarray}
B_{435}  & > & 27.5,\\
V_{606}  & > & 27.6, {\rm and}\\
iz-{\NB} & > & 0.7
\end{eqnarray}
where $iz$ is the continuum magnitude at $\lambda = 8150$ \AA~, estimated 
by linear interpolation between 
$i_{775}$ and $z_{850}$ flux densities 
($f_{iz}= 0.63 f_{i_{775}} + 0.37 f_{z_{850}}$).
The first two criteria ensure that the objects at $z\approx 5.7$ are
undetected
(at 2 $\sigma$ noise level) in $B_{435}$ and $V_{606}$ while, the third
condition allows selection of LAEs with a rest-frame Lyman $\alpha$ equivalent 
width $\geq 17$ \AA. This is illustrated in 
Figure \ref{nbcm} where the LAE candidates are identified on 
$iz -  \NB$ {\it vs.} \NB. color-magnitude diagram. A total of 10 (GOODS-N)
and 4 (GOODS-S) LAEs at a targeted redshift of $z\sim 5.7$ are found to 
satisfy the above criteria. Their
\NB ~ and broad-band ACS images are shown in Figure \ref{thumN} (GOODS-N)
and \ref{thumS} (GOODS-S). Their coordinates and photometric properties are 
summarized in Table \ref{tab:LAE}.

It is important to estimate the fraction of contaminants in our LAE survey. 
From spectroscopic observations of a sample of 23 LAE candidates at
$z\sim 5.7$, identified in a narrow-band survey,
 Hu et al. (2004) found that 4 are interlopers
 [two are [O {\sc ii}] and H$\alpha$ emitters, one is a red star
 ($\NB -z^\prime > 0.2$) and one a source with no discernible
 spectral features]\footnote{
         Hu et al. (2004) used the following criteria for selecting their LAE candidates;
         (a) $\NB < 25.05$, (b) $I_{\rm C} - \NB > 0.7$, (c) undetected in the $B$ and $V$ images,
         and (d) $r^\prime - z^\prime > 1.8$.
}. 
Since none of our LAE candidates have red $\NB - z_{850}$ colors,
 similar to what expected for stars, 
 our LAE survey is expected to be free from contamination by red stars.
This, combined with the fact that the $B$- and $V$-band data here are deeper 
than those of Hu et al. (2004) by $\sim 1$ magnitude,  
we expect a contamination rate of $<$14\% for our sample of
LAEs here.

\section{Results }
\subsection{Number Density and Lyman $\alpha$ Luminosities of LAE Candidates}

\begin{figure}
\epsscale{1.0}
\plottwo{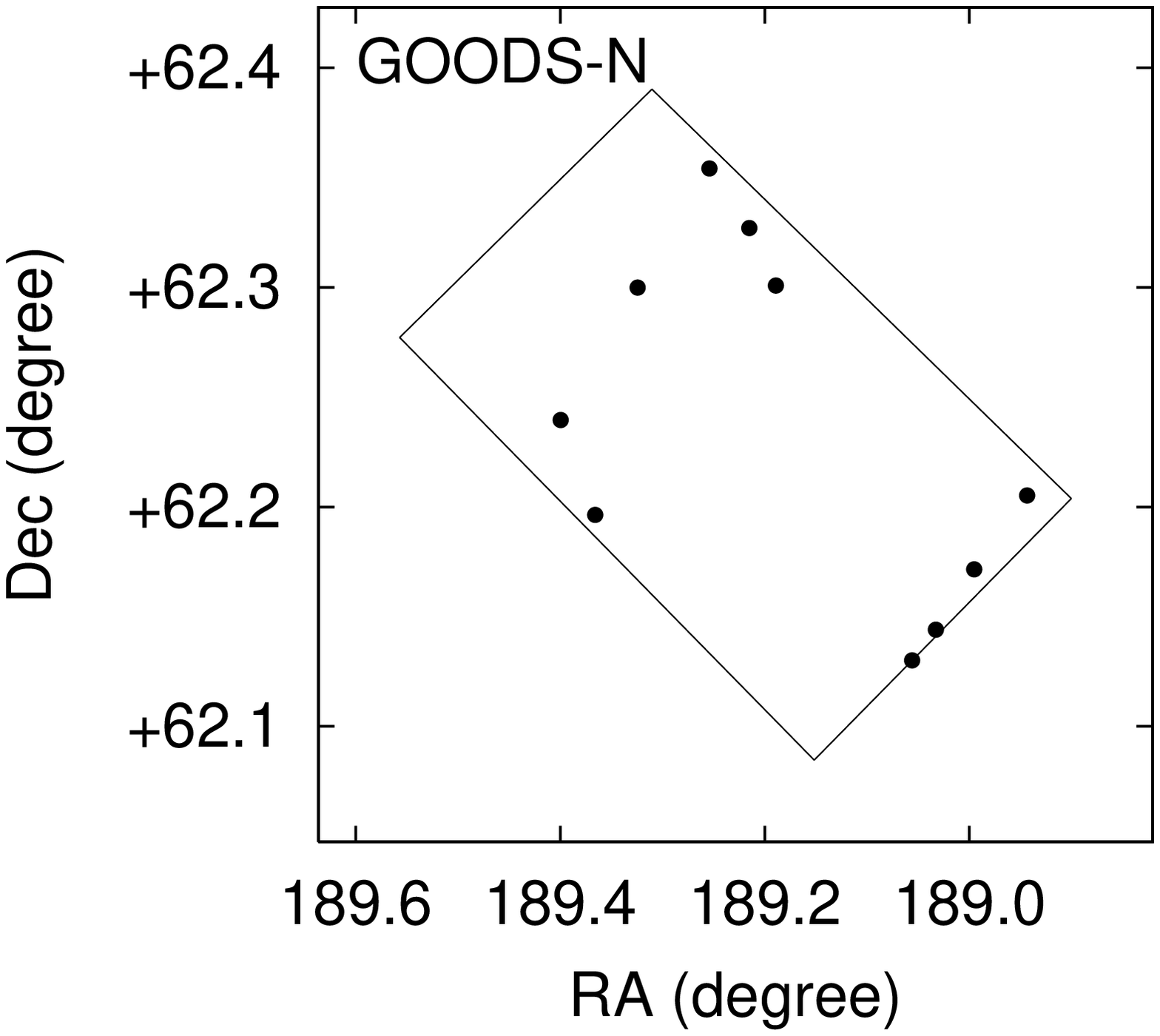}{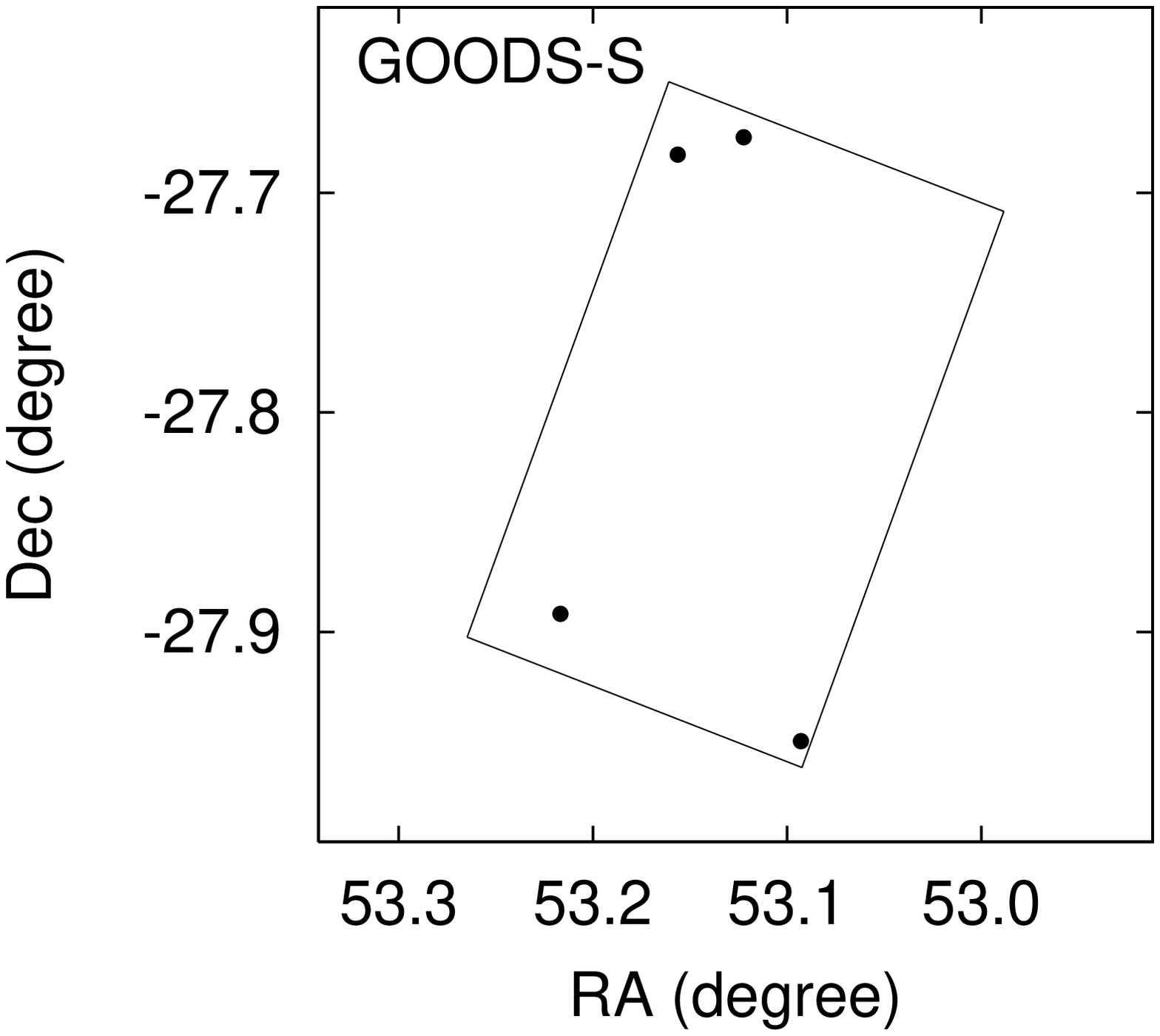}
\caption{Spatial distributions of our LAE candidates in GOODS-N (left) and -S (right).
\label{map}}
\end{figure}

Using the estimated survey volume for each field, 
$4.0 \times 10^{4} $ Mpc$^{3}$, and the number of LAE candidates
found at $z \approx 5.7$, we estimate the LAE space density of
$n($LAE$) \simeq 2.5 \times 10^{-4}$Mpc$^{-3}  $ and
$\simeq 1.0 \times 10^{-4}$Mpc$^{-3}$ for GOODS-N and
GOODS-S respectively. The total space density in the combined GOODS fields
is then $n($LAE$)\simeq 1.7 \times 10^{-4}$Mpc$^{-3}$.
The space density of LAEs in the GOODS-S is lower by a factor of 2.5 than 
that in GOODS-N. Spatial distribution of the LAE candidates are shown in Figure \ref{map}.

\begin{figure}
\epsscale{0.5}
\plotone{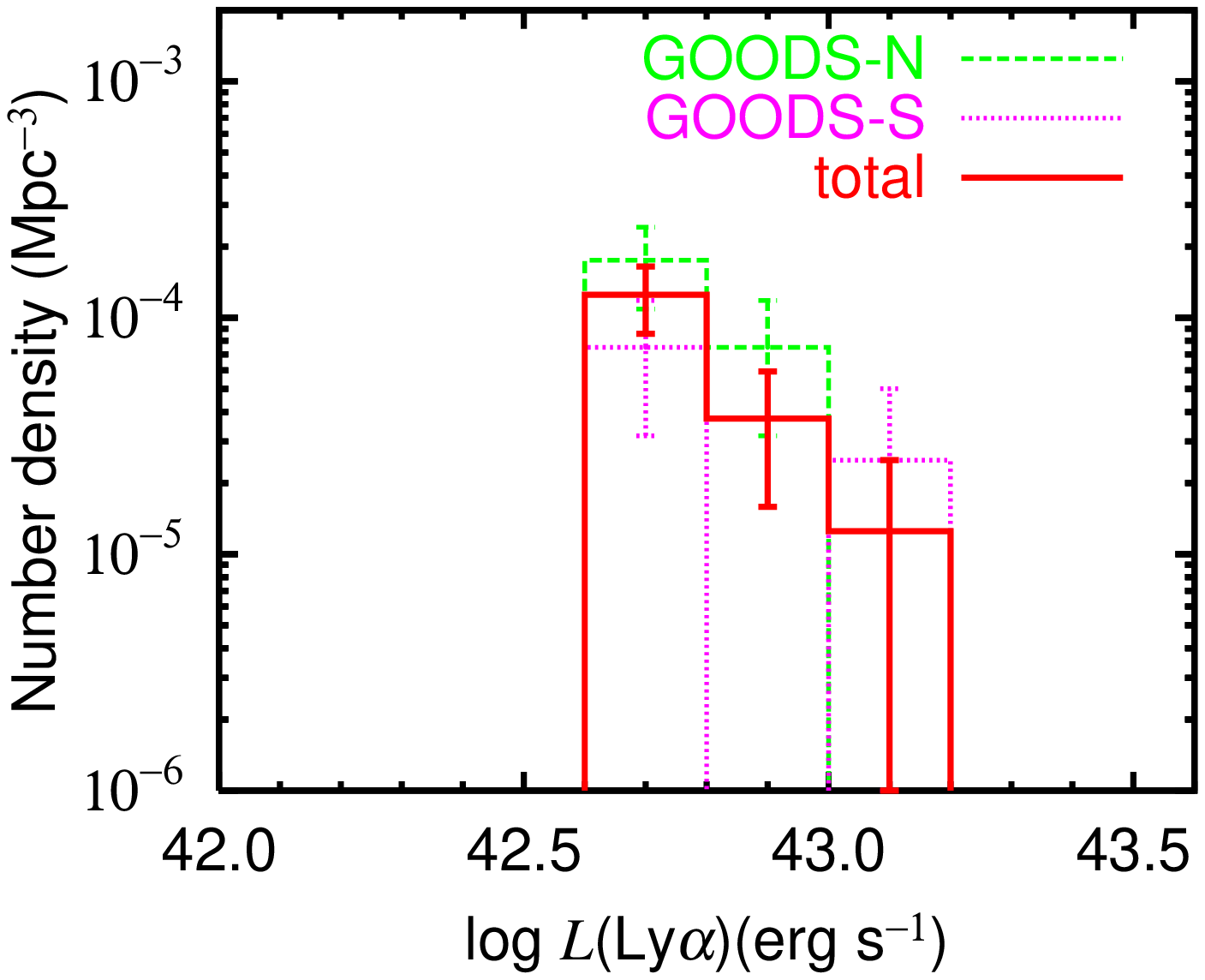}
\plotone{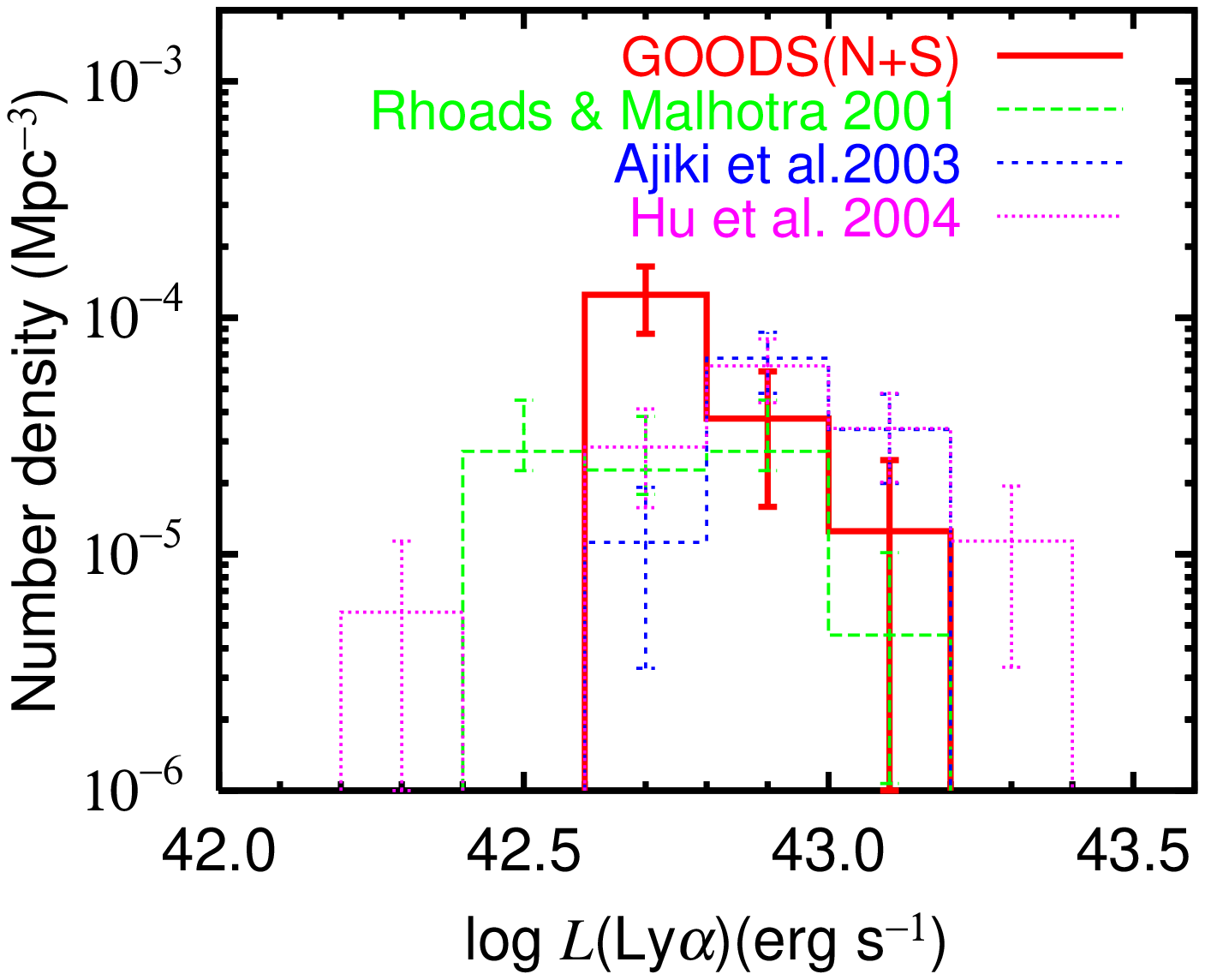}
\caption{Upper panel: Distribution of Lyman $\alpha$ luminosity of our LAE candidates.
         Lower panel: Our results is compared with those from 
         Rhoads \& Malhotra (2001), Ajiki et al. (2003) and Hu et al. (2004).
\label{ll}}
\end{figure}

In Table \ref{tab:LLAE}, we list the Lyman $\alpha$ luminosities for
our sample of LAE candidates, assuming them to be at $z=5.70$, with 
the Lyman $\alpha$ shifted to the central wavelength of the \NB ~ filter. 
The Lyman $\alpha$ luminosities range from
$\approx 4.2 \times 10^{42} $ ergs s$^{-1}$
to $\approx 1.2 \times 10^{43} $ ergs s$^{-1}$.  
These values correspond to $L_*$ -- $3 L_*$ of the Ly$\alpha$ luminosity
 function by Ajiki et al. (2003; $L_*=4.1 \pm 0.4 \times 10^{42}$ ergs s$^{-1}$).
In Figure \ref{ll}, we show the distribution of Lyman $\alpha$ luminosities for
our 14 LAE candidates (upper panel).
We compare our results with those of the previous LAE surveys at $z=5.7$
(Rhoads \& Malhotra 2001; Ajiki et al. 2003; Hu et al. 2004) (lower panel 
in Figure \ref{ll}).  It is clear that the number density of our sample of 
 LAEs with Lyman $\alpha$ luminosities of $\approx 5 \times 10^{42} $ 
ergs s$^{-1}$, is higher (by a factor of $\sim 5$) than those of the other 
surveys. 
This seems attributed to the relatively deeper completeness limit for our survey
[$L_{\rm lim}$(Ly$\alpha$)$ \simeq 4.8 \times 10^{42} $ ergs s$^{-1}$
corresponding to $\NB=25.4$] compared to the previous studies, including; 
Rhoads \& Malhotra [2001; 
$L_{\rm lim}$(Ly$\alpha$)$ \simeq 5.3 \times 10^{42} $ ergs s$^{-1}$], 
Ajiki et al. [2003;
$L_{\rm lim}$(Ly$\alpha$)$ \simeq 7.0 \times 10^{42} $ ergs s$^{-1}$],
and
Hu et al. [2004; 
$L_{\rm lim}$(Ly$\alpha$)$ \simeq 6.6 \times 10^{42} $ ergs s$^{-1}$]. 
The extra depth enables our survey to include more LAEs with the 
Lyman $\alpha$ luminosities
$\sim 4 -- 6 \times 10^{42} $ ergs s$^{-1}$ than the previous surveys.
We also note that, for luminosity [$L$(Ly$\alpha$)$ > 7 \times 10^{42} $ ergs s$^{-1}$], 
 the number density of our combined (GOODS-N and -S) sample
 is higher than that of Rhoads \& Malhotra (2001) by a factor of $\approx 2$ while
 lower than those of Ajiki et al.(2003) and Hu et al. (2004) by a similar
factor.  

\subsection{Star Formation Rates}

Using the LAE luminosities, we now estimate their corresponding star formation 
rates (SFRs), using the relation 
\begin{equation}
SFR({\rm Ly}\alpha)  = 9.1 \times 10^{-43} L({\rm Ly}\alpha) ~ M_\odot ~ {\rm yr}^{-1},
\end{equation}
(Kennicutt 1998; Brocklehurst 1971). Here 
$L($Ly$\alpha)$ is in units of ergs s$^{-1}$ and we assume a
Salpeter initial mass function with ($m_{\rm lower}$, $m_{\rm upper}$)
= (0.1 $M_\odot$, 100 $M_\odot$).
The results are given in the third column of Table \ref{tab:LLAE}.
We note that the SFR derived here can be underestimated due to the effect of
absorption by H {\sc i} gas both in the host galaxies and in 
the intergalactic space.
The SFRs range from 4 to 11 $ M_\odot$ yr$^{-1}$
with a median of 5.2 $M_\odot$ yr$^{-1}$.  
The estimated SFRs here are comparable to those of LAEs at 
$z \simeq$ 5.7 -- 6.6 (e.g., Ajiki et al. 2003; Taniguchi et al. 2005).

\begin{figure}
\epsscale{1.0}
\plottwo{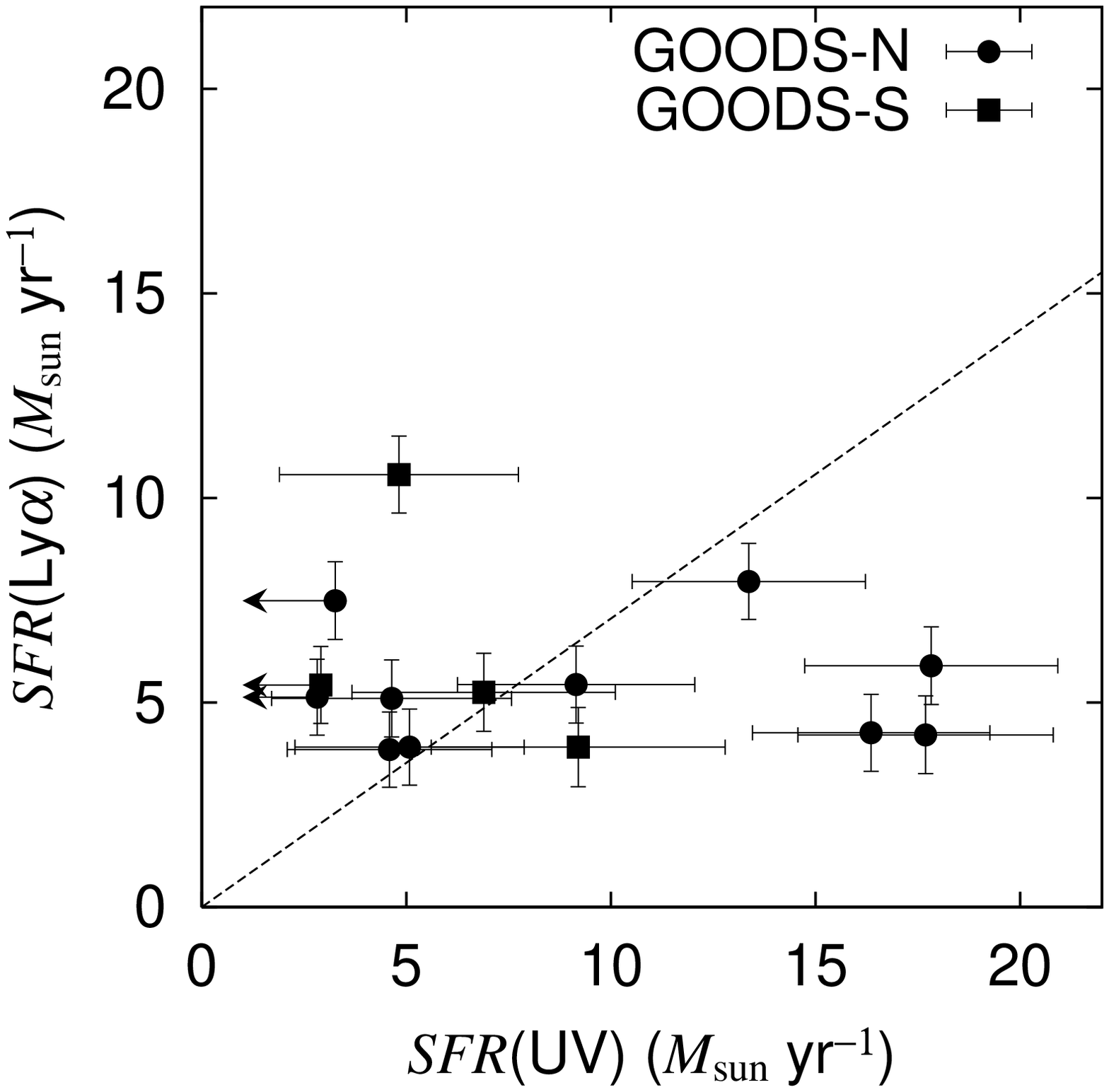}{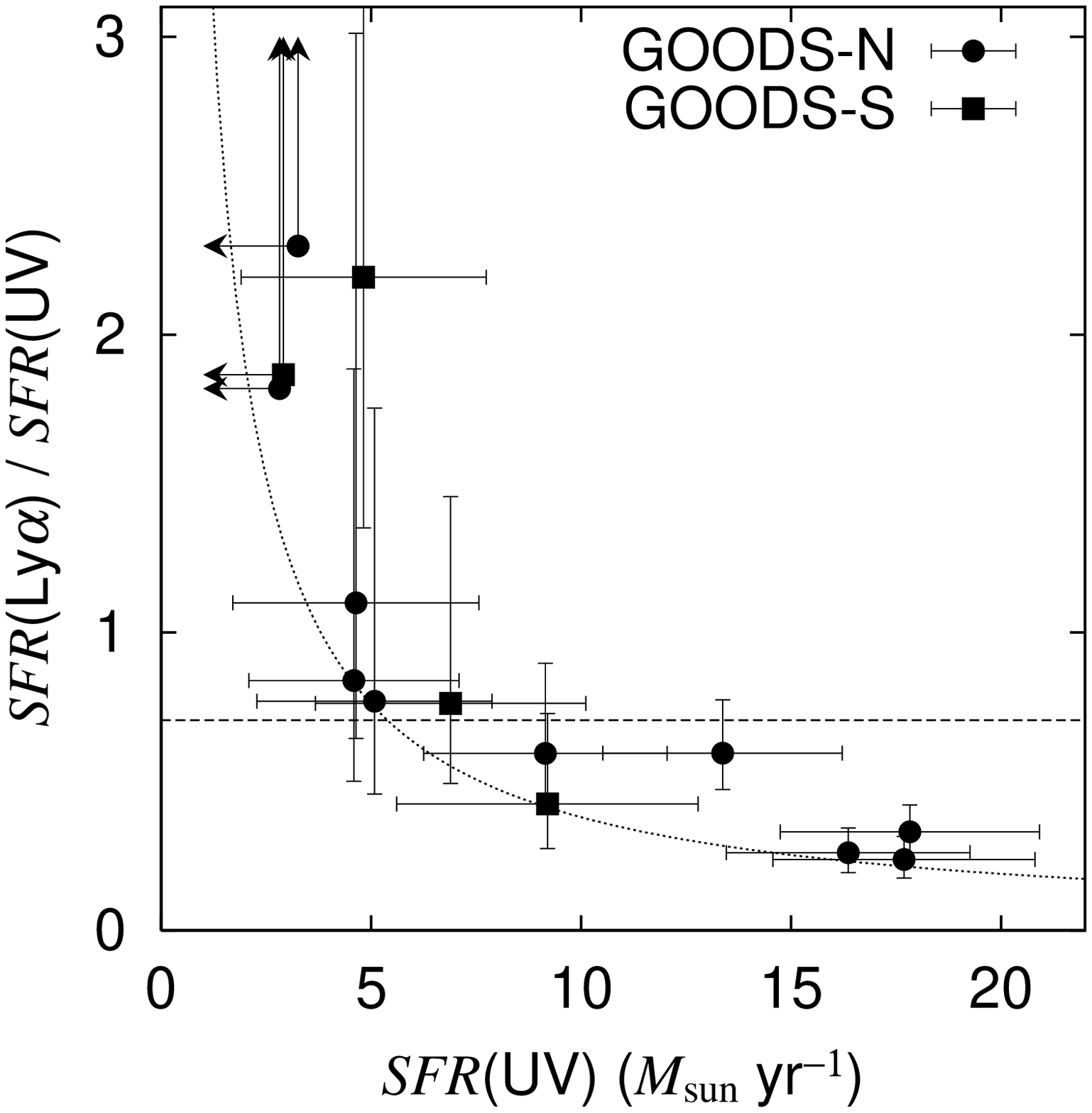}
\caption{Comparison between $SFR$(Ly$\alpha$) and $SFR$(UV) (left panel)
         and $SFR$(Ly$\alpha$) / $SFR$(UV) and $SFR$(UV) (right panel) for 14 LAE candidates. 
         The dashed lines in the both panels show $SFR($Ly$\alpha) /SFR($UV$) =0.71$.
         The dotted curve in the right panel show detection limit of our survey
          corresponding to $SFR($Ly$\alpha$) $= 3.8  M_\odot$ yr$^{-1}$.
\label{sfr}}
\end{figure}

It is instructive to examine whether the SFRs derived from  
 Lyman $\alpha$ luminosities here are consistent with those derived from
 the UV continuum luminosity from the broad-band data in the GOODS. We
convert the observed $z_{850}$ magnitudes to
 UV continuum luminosities at $\lambda = 1360$ \AA.
Using the relation (Kennicutt 1998; see also Madau et al. 1998),
\begin{equation}
\label{UVtoSFR}
SFR({\rm UV}) = 1.4 \times 10^{-28}L_{\nu}
~~ M_{\odot} ~ {\rm yr}^{-1},
\end{equation}
 where $L_\nu$ is in units of ergs s$^{-1}$ Hz$^{-1}$. The estimated SFRs are 
summarized in Table \ref{tab:LLAE}.
Comparison between the SFRs for individual sources, estimated from 
 $SFR$(Ly$\alpha$) and $SFR$(UV) (see Figure \ref{sfr}),   
 shows that, on average, $SFR$(UV) is relatively higher than 
$SFR$(Ly$\alpha$) for most of the LAE candidates.    
We also measure the average SFR ratios,  
$SFR_{\rm total}$(Ly$\alpha$) / $SFR_{\rm total}$(UV)$ = 0.71$,
where $SFR_{\rm total}$(Ly$\alpha$) and $SFR_{\rm total}$(UV)
are, respectively, the sum of $SFR$(Ly$\alpha$) and $SFR$(UV) 
of all our LAE candidates. 
This ratio is close to that obtained by Ajiki et al. (2003)
for their sample of LAEs at $z \approx 5.7$.
The lower value of $SFR$(Ly$\alpha$) can be attributed to the effect of
 the dust extinction and scattering by the intergalactic medium.
However, some of our LAE candidates have $SFR$(Ly$\alpha$) / $SFR$(UV) $ > 1$.
 These are likely to be in a very early phase ($<10^8$ yr) of star formation
activity, in which,  $SFR$(UV) values are underestimated (Schaerer 2000; 
 see also Nagao et al. 2004, 2005). 
As illustrated in the right panel of Figure \ref{sfr}, objects 
with small $SFR$(UV) have higher $SFR$(Ly$\alpha$) / $SFR$(UV) ratios.  
This, in part, is explained by selection effects in 
 the \NB -limited sample (i.e., our survey cannot find objects
  below the dotted curve in the right panel of Figure \ref{sfr}).
However, all objects with $SFR$(Ly$\alpha$) / $SFR$(UV) $ > 1$
 (i.e., the age of $<10^8$ yr) have $SFR$(UV) $\lesssim 5$.
This implies that the UV continuum emission of very young 
  star-forming galaxies are too faint to be found in typical LBG surveys.
The narrow-band survey provides a very 
 efficient method to investigate such very young star-forming galaxies.

\section{Summary}

We present results from a \NB~ survey in the two GOODS fields,
 aiming for LAE galaxies in the redshift range $z=5.65-5.75$. 
We find a total of 14 LAE candidates satisfying our selection criteria. 
The space density of LAEs with $L$(Ly$\alpha$) $\approx 5 \times 10^{42}$
 ergs s$^{-1}$, found here ($1.0\times 10^{-4}$ Mpc$^{-3}$),
 is a factor of $\sim 5$ higher than other similar surveys
 (Rhoads \& Malhotra 2001; Ajiki et al. 2003; Hu et al. 2004),
 due to relatively deeper flux limit in the present survey.
Using the Ly$\alpha$ luminosities, we estimate the SFRs for LAEs,
 $SFR$(Ly$\alpha$) to be in the range $4-11$ M$_\odot$ yr$^{-1}$.
Overall, this is smaller than the SFR estimated from rest-frame $UV$ light
 [$SFR$(Ly$\alpha$)/$SFR$(UV)$ = 0.71$], due to extinction in Ly$\alpha$ flux.
We interpret a larger $SFR$(Ly$\alpha$) value
 [i.e. $SFR$(Ly$\alpha$) / $SFR$(UV)$ > 1 $], observed for a few LAEs as
 due to a very early phase of star formation ($< 10^8$ yr).

\vspace{3ex}

We would like to thank both the Subaru and HST staff for their invaluable help.
Data reduction/analysis was in part carried out on "sb" computer system
operated by the Astronomical Data Analysis Center (ADAC) and Subaru Telescope of
the National Astronomical Observatory of Japan.
We also thank an anonymous referee for his/her useful comments and suggestions.
This work was financially supported in part by the Ministry
of Education, Culture, Sports, Science, and Technology (Nos. 10044052 
and 10304013), and by JSPS (15340059 and 17253001). MA, SSS, and TN
are JSPS fellows.

%
%\newpage
%-------------------------------------------------------------------------

%------------------------------
% Tabl 1: LAE table
%------------------------------

\begin{deluxetable}{rcclllll}
\tabletypesize{\scriptsize}
\tablenum{1}
\tablecaption{the LAE candidates at $\approx 5.7$ in GOODS-N and GOODS-S 
fields. \tablenotemark{a} \label{tab:LAE}}
\tablewidth{0pt}
\tablehead{
\colhead{No.} &
\colhead{$\alpha$(J2000)} &
\colhead{$\delta$(J2000)} &
\colhead{$B_{435}$\tablenotemark{a}} &
\colhead{$V_{606}$\tablenotemark{a}} &
\colhead{$i_{775}$\tablenotemark{a}} &
\colhead{$z_{850}$\tablenotemark{a}} &
\colhead{\NB\tablenotemark{a}} \\
\colhead{} &
\colhead{h ~~ m ~~ s} &
\colhead{$^\circ$ ~~ $^\prime$ ~~ $^{\prime\prime}$}  &
\colhead{}&
\colhead{}&
\colhead{}&
\colhead{}&
\colhead{}
}
\startdata
\multicolumn{8}{c}{GOODS-N}\\
\tableline
 1 & 12 35 46.4 & +62 12 19 & 29.26  & 99     & 26.91  & 26.91  & 25.12  \\
 2 & 12 35 58.9 & +62 10 18 & 27.27  & 27.87  & 26.65  & 26.81  & 25.37  \\
 3 & 12 36 07.9 & +62 08 39 & 27.56  & 28.48  & 26.17  & 28.09  & 24.79  \\
 4 & 12 36 13.4 & +62 07 48 & 29.51  & 34.20  & 26.30  & 25.44  & 24.75  \\
 5 & 12 36 45.4 & +62 18 03 & 29.07  & 99     & 28.35  & 26.92  & 25.40  \\
 6 & 12 36 51.6 & +62 19 37 & 99     & 28.53  & 26.82  & 25.76  & 24.56  \\
 7 & 12 37 01.0 & +62 21 15 & 99     & 29.13  & 26.10  & 25.54  & 25.03  \\
 8 & 12 37 17.9 & +62 17 59 & 27.61  & 31.58  & 26.74  & 26.17  & 24.97  \\
 9 & 12 37 27.8 & +62 11 47 & 99     & 99     & 27.43  & 99     & 25.23  \\
10 & 12 37 35.9 & +62 14 22 & 27.27  & 29.08  & 26.22  & 25.45  & 25.01  \\
\tableline
\multicolumn{8}{c}{GOODS-S}\\
\tableline
11 & 03 32 22.3 & -27 56 59 & 27.34  & 27.92  & 28.40  & 26.16  & 25.26  \\
12 & 03 32 29.4 & -27 40 29 & 27.65  & 27.24  & 25.89  & 26.48  & 25.05  \\
13 & 03 32 37.5 & -27 40 57 & 30.36  & 28.93  & 26.90  & 26.87  & 24.39  \\
14 & 03 32 52.0 & -27 53 31 & 28.14  & 29.06  & 26.99  & 99     & 25.17  \\
\enddata
\tablenotetext{a}{AB magnitude in a $2^{\prime \prime}$ diameter.
 An entry of ``99'' indicates that no excess flux was measured}
\end{deluxetable}

%-------------------------------------------------------------------------
%                 Table 2
%-------------------------------------------------------------------------
\begin{deluxetable}{rrrrrr}
\tablenum{2}
\tabletypesize{\scriptsize}
\tablecaption{Lyman $\alpha$ luminosities and star formation rates for the
LAE candidates at $z \approx 5.7$\label{tab:LLAE}}
\tablewidth{0pt}
\tablehead{
\colhead{No.} &
\colhead{$L({\rm Ly}\alpha)$\tablenotemark{a}} &
\colhead{$SFR$(Ly$\alpha$)\tablenotemark{b}}  &
\colhead{$L_{1360}$\tablenotemark{c}} &
\colhead{$SFR$(UV)\tablenotemark{d}} &
\colhead{$SFR({\rm Ly\alpha})/SFR({\rm UV})$} \\
\colhead{} &
\colhead{($10^{42}$ ergs s$^{-1}$)} &
\colhead{($M_\odot$ yr$^{-1}$)} &
\colhead{($10^{28}$ ergs s$^{-1}$ Hz$^{-1}$)} &
\colhead{($M_\odot$ yr$^{-1}$)} &
\colhead{} \\
}
\startdata
 1 & 5.6  & 5.0  &    3.3  &  4.6  & 1.09  \\
 2 & 4.3  & 3.9  &    3.6  &  5.1  & 0.76  \\
 3 & 8.2  & 7.4  & $<$2.0 & $<$2.8 & $>$2.65  \\
 4 & 6.5  & 5.8  &   12.7  & 17.8  & 0.33  \\
 5 & 4.2  & 3.8  &    3.3  &  4.6  & 0.83  \\
 6 & 8.7  & 7.9  &    9.6  & 13.4  & 0.59  \\
 7 & 4.7  & 4.2  &   11.7  & 16.4  & 0.26  \\
 8 & 6.0  & 5.4  &    6.5  &  9.2  & 0.59  \\
 9 & 5.6  & 5.1  & $<$2.0 & $<$2.8 & $>$1.81 \\
10 & 4.6  & 4.2  &   12.6  & 17.7  & 0.24  \\
\tableline
11 & 4.3  & 3.9  &    6.6  &  9.2  & 0.42  \\
12 & 5.8  & 5.2  &    4.9  &  6.9  & 0.75  \\
13 & 11.6 & 10.5 &    3.4  &  4.8  & 2.17  \\
14 & 6.0  & 5.4  & $<$2.0 & $<$2.8 & $>$1.92 \\
\enddata
\tablenotetext{a}{ $\sigma \approx 0.1 \times 10^{43}$ ergs
s$^{-1}$.}
\tablenotetext{b}{ $\sigma \approx 1~ M_\odot$ yr$^{-1}$.}
\tablenotetext{c}{The UV continuum luminosity at $\lambda=1360$ \AA.
             $\sigma \approx 2.0 \times 10^{28}$ ergs s$^{-1}$ Hz$^{-1}$.}
\tablenotetext{d}{ $\sigma \approx 2.8 M_\odot$ yr$^{-1}$.}
\end{deluxetable}

%---------------------------------------------------------------------
% Figure: cm
%---------------------------------------------------------------------
\clearpage

%---------------------------------------------------------------------
% Figure: thumbnail N
%---------------------------------------------------------------------
\begin{figure}
\epsscale{1.0}
\plotone{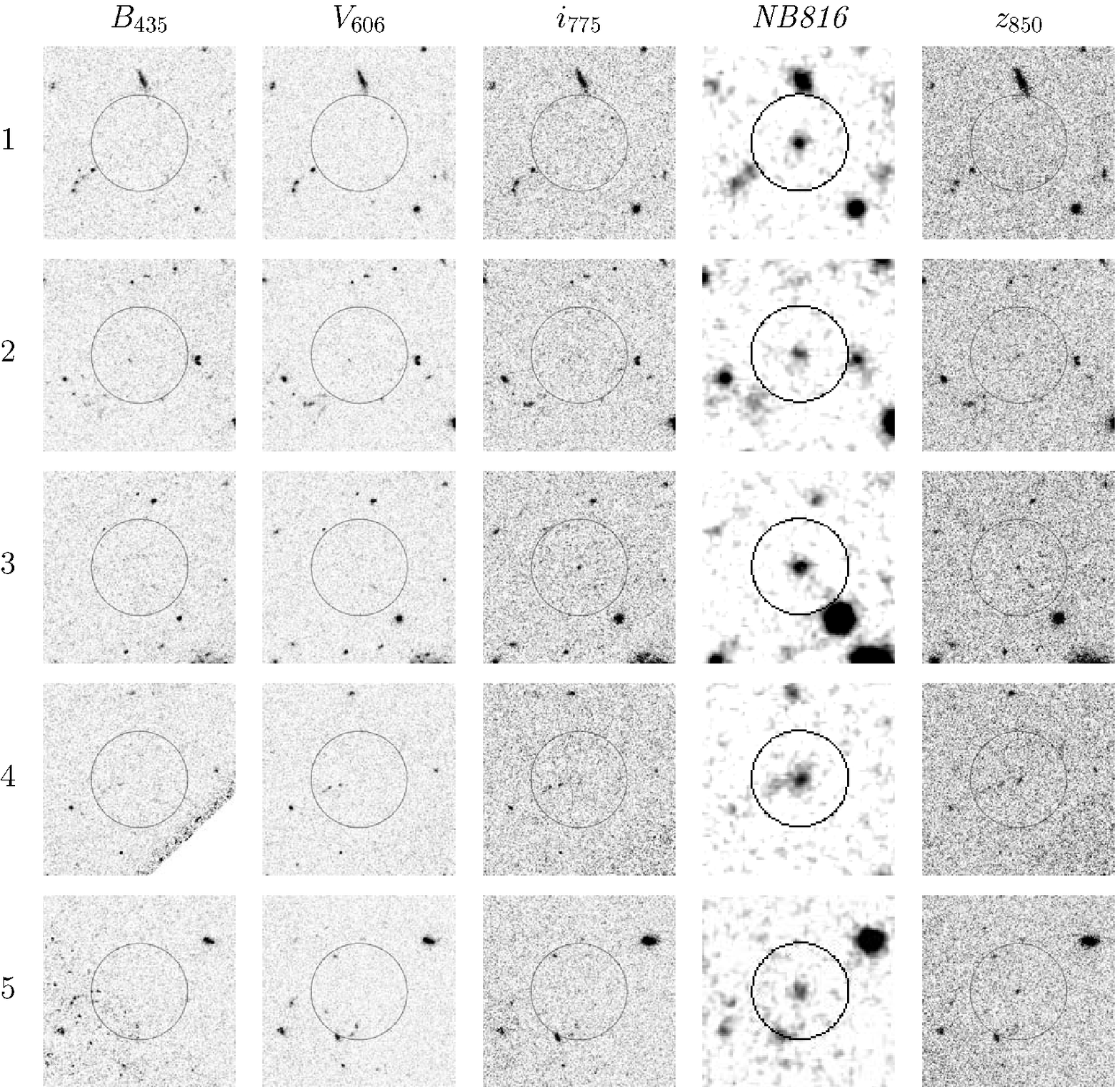}
\caption{Broad-band and \NB ~ images of our
          LAE candidates at $z \approx 5.7$ in GOODS-N.
         Each box is $12^{\prime \prime}$ on a side
         (north is up and east is left).
         Each circle has $3^{\prime \prime}$ radius.\label{thumN}
}
\end{figure}

\begin{figure}
\epsscale{1.0}
\figurenum{\ref{thumN}}
\plotone{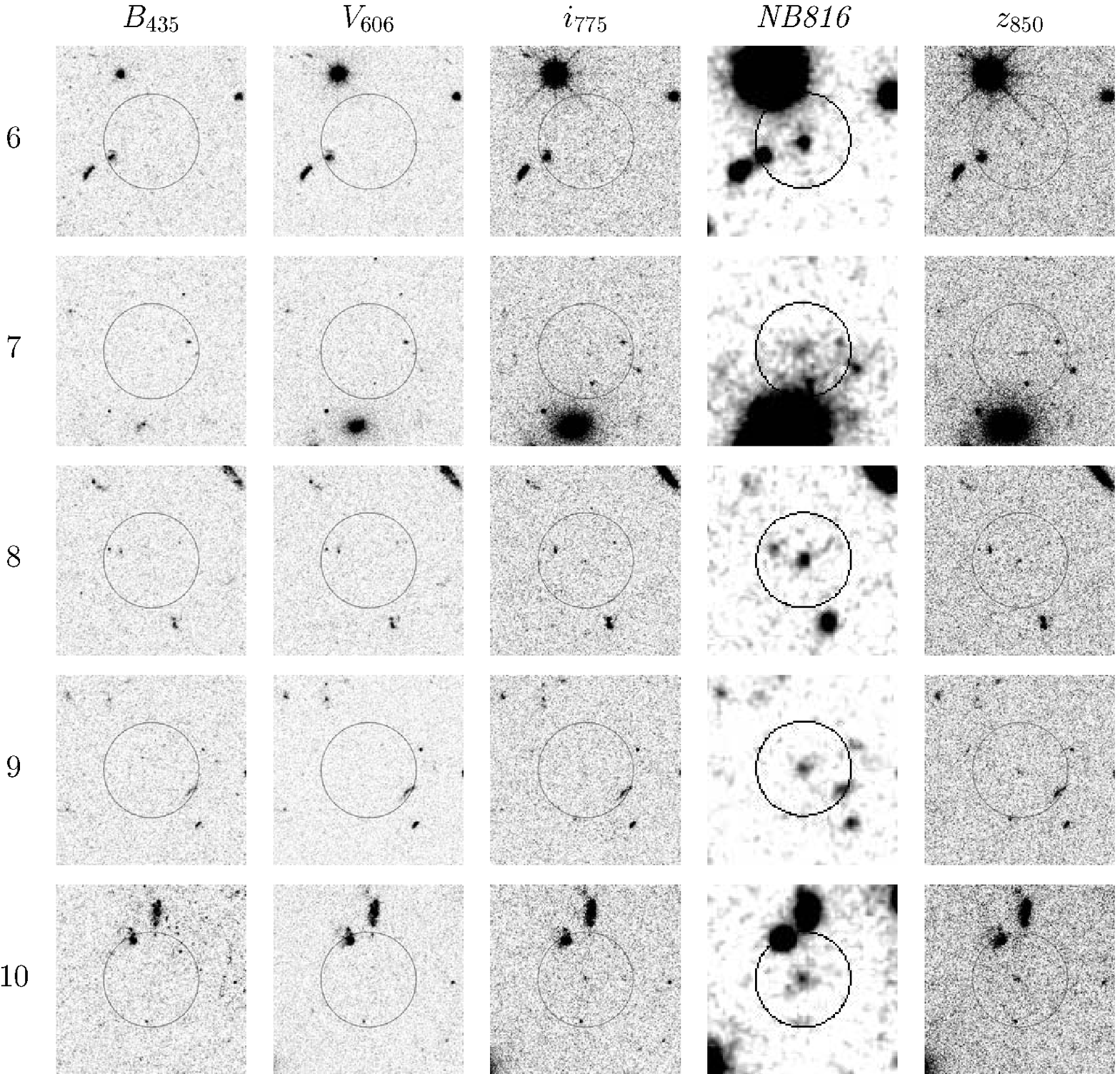}
\caption{continued.}
\end{figure}

%---------------------------------------------------------------------
% Figure: thumbnail S
%---------------------------------------------------------------------

%---------------------------------------------------------------------
% Figure: spatial distributions
%---------------------------------------------------------------------

%---------------------------------------------------------------------
% Figure Lyman alpha
%---------------------------------------------------------------------
%\clearpage

%---------------------------------------------------------------------
% Figure sfr
%---------------------------------------------------------------------

\end{document}